\documentclass[12pt]{article}
\usepackage{mathrsfs}
\usepackage{graphicx}
\usepackage{amsmath}
\usepackage{amssymb}
\usepackage{caption2}
\begin{document}
\newcommand  {\ba} {\begin{eqnarray}}
\newcommand  {\be} {\begin{equation}}
\newcommand  {\ea} {\end{eqnarray}}
\newcommand  {\ee} {\end{equation}}
\renewcommand{\thefootnote}{\fnsymbol{footnote}}
\renewcommand{\figurename}{Figure.}
\renewcommand{\captionlabeldelim}{.~}

\vspace*{1cm}
\begin{center}
 {\Large\textbf{A Model of Four Generation Fermions and Cold Dark Matter and Matter-Antimatter Asymmetry}}

\vspace{1cm}
 \textbf{Wei-Min Yang}

\vspace{0.4cm}
 \emph{Department of Modern Physics, University of Science and Technology of China, Hefei 230026, P. R. China}

\vspace{0.2cm}
 \emph{E-mail: wmyang@ustc.edu.cn}
\end{center}

\vspace{1cm}
 \noindent\textbf{Abstract}: I suggest a practical particle model as an extension to the standard model. The model has a TeV scale $U(1)_{B-L}$ symmetry and it contains the fourth generation fermions with the TeV scale masses, in which including a cold dark matter neutrino. The model can completely account for the fermion flavor puzzles, the cold dark matter, and the matter-antimatter asymmetry through the leptogenesis. In particular, it is quite feasible and promising to test the model in future experiments.

\vspace{1cm}

\vspace{0.3cm}

\newpage
 \noindent\textbf{I. Introduction}

\vspace{0.3cm}
 The precise tests for the electroweak scale physics have established plenty of knowledge about the elementary particles \cite{1,2}. The standard model (SM) has been evidenced to be indeed a very successful theory at the electroweak energy scale. Nevertheless, there are a lot of the issues in the flavor physics and universe observations for which the SM is not able to account \cite{3}. During the past more decades a series of experiment results of $B$ physics and neutrino physics have told us a great deal of the detailed information of flavor physics \cite{4}. What are paid more attention are some facts in the following. Firstly, the fermion mass spectrum emerges a very large hierarchy. The quark and charged lepton masses range from one MeV to one hundred GeV or so \cite{1}, while the neutrino masses are only at the Sub-eV level \cite{5}. Secondly, the quark flavor mixing pattern is distinctly different from the lepton one. The former are only three small mixing angles \cite{6}, whereas the latter has bi-large mixing angles and a bit small but non-zero $\theta_{13}$ \cite{7}. In addition, the $CP$ violation in the quark sector has been verified to be non-zero but very small \cite{8}, however, it is in suspense whether the $CP$ violation in the lepton sector vanishes or not \cite{9}. Thirdly, that the light neutrino nature is Majorana or Dirac particle has to be further identified by the experiments such as $0\nu\beta\beta$ \cite{10}. On the other hand, the astrophysics observations and researches lead to some impressive puzzles in the universe, in particular, the genesis of the matter-antimatter asymmetry and the original nature of cold dark matter \cite{11}. Finally, a very important and unsolved problem is whether flavor physics are truly connected with the baryon asymmetry and/or cold dark matter in the universe or not? All the problems have a great significance for both particle physics and the early universe \cite{12}, so they always attract great attention in the experiment and theory fields.

 The researches for the above-mentioned problems have motivated many new theories beyond the SM. The various theoretical approaches and models have been proposed to solve the intractable issues \cite{13}. For instance, the Froggatt-Nielsen mechanism with the $U(1)$ family symmetry can account for mass hierarchy \cite{14}, the discrete family group $A_{4}$ can lead to the tri-bimaximal mixing structure of the lepton mixing matrix \cite{15}, the non-Abelian continuous group $SU(3)$ is introduced to explain the neutrino mixing \cite{16}, the model with the $SO(3)$ family group can accommodate the experimental data of the quarks and leptons by the fewer parameters \cite{17}, some grand unification models based on the $SO(10)$ symmetry group can also give some reasonable interpretations for fermion masses and flavor mixings \cite{18}. In addition, some suggestions of the baryon asymmetry and cold dark matter are very constructive \cite{19}. Although these theories are successful in explaining some specific aspects of the above problems, it seems very difficult for them to solve the whole problems all together. Indeed it is especially hard for a model construction to keep the principle of the simplicity, economy and the less number of parameters, otherwise the theory will be excessive complexity and incredible. On all accounts, a better theory beyond the SM has to be confronted with the integration of particle physics and the early universe, in other words, it should simultaneously account for flavor physics, the baryon asymmetry and cold dark matter. Of course, it is still a large challenge for theoretical particle physicists to find a desirable theory to uncover these mysteries of the universe \cite{20}.

 In this work, I construct a practical model to integrate three parts of the fermion flavor, cold dark matter and leptogenesis. The model has the local gauge symmetries $SU(3)_{C}\otimes SU(2)_{L}\otimes U(1)_{X}\otimes U(1)_{B-L}$. The subgroups $U(1)_{X}\otimes U(1)_{B-L}$ only appear above the TeV scale, and then are spontaneously broken to the supercharge subgroup $U(1)_{Y}$ below that scale. Besides the SM particles, some new non-SM particles are introduced into the model. They are the fourth generation quarks and leptons, two vector gauge fields related to $U(1)_{X}\otimes U(1)_{B-L}$, and three scalar fields, which are respectively one neutral singlet, one charged singlet and one symmetric triplet under $SU(2)_{L}$. The $U(1)_{X}\otimes U(1)_{B-L}$ breaking is accomplished by the neutral singlet scalar developing a non-vanishing vacuum expectation value (VEV) at the TeV scale. This breaking leads that one of the two new gauge fields obtains a TeV scale masse by the Higgs mechanism, in the meantime, the fourth generation neutrino is given rise to a TeV scale mass. The fourth generation neutrino has some unique natures in the model, which ensure that it is a cold dark matter particle. All of the SM particle masses are generated after the electroweak breaking. The triplet scalar field takes part in the symmetry breakings but it develops only a tiny VEV by virtue of its very heavy mass. This is a real source of the tiny masses of the light neutrinos. The leptogenesis is implemented by the charged singlet scalar decaying into a SM charged lepton and a cold dark matter neutrino. The decay process simultaneously satisfies the three conditions of the $B-L$ violation, $CP$ violation and being out-of-equilibrium. This mechanism can naturally generate the $B-L$ asymmetry, subsequently it is converted into the baryon asymmetry through the electroweak sphaleron process \cite{21}. In the model, the flavor physics is intimately associated with the cold dark matter and the matter-antimatter asymmetry in the universe. The model can completely accommodate and fit all the current experimental data of the fermion masses and mixings, the cold dark matter and the baryon asymmetry, furthermore, it also predicts some interesting results. Finally, the model is feasible and promising to be tested in future experiments. I give some methods of searching the non-SM particles of the model in the experiments at the LHC \cite{22}.

 The remainder of this paper is organized as follows. In Section II I outline the model. In Sec. III, the model symmetry breakings, the particle masses and mixings are introduced. In Sec. IV, the leptogenesis and cold dark matter are discussed. In Sec. V, I give the detailed numerical results. Sec. VI is devoted to conclusions.

\vspace{1cm}
 \noindent\textbf{II. Model}

\vspace{0.3cm}
 The local gauge symmetries of the model are the direct product groups of $SU(3)_{C}\otimes SU(2)_{L}\otimes U(1)_{X}\otimes U(1)_{B-L}$. The subgroup symmetry of $U(1)_{X}\otimes U(1)_{B-L}$ only appears above the TeV scale, and it will be broken to the supercharge subgroup $U(1)_{Y}$ below that scale. The model particle contents and their gauge quantum numbers are in detail listed in the following,
\begin{alignat}{1}
 & G_{\mu}^{a}(8,1,0,0)\,,\hspace{0.5cm} W_{\mu}^{i}(1,3,0,0)\,,\hspace{0.5cm} X_{\mu}(1,1,1,0)\,,\hspace{0.5cm}
   Y_{\mu}(1,1,0,1)\,,\nonumber\\
 & Q_{i}(3,2,0,\frac{1}{3})\,,\hspace{0.5cm} [u_{iR},u_{4R},u_{4L}](3,1,1,\frac{1}{3})\,,\hspace{0.5cm}
   [d_{iR},d_{4R},d_{4L}](3,1,-1,\frac{1}{3})\,,\nonumber\\
 & L_{i}(1,2,0,-1)\,,\hspace{0.5cm} [e_{iR},e_{4R},e_{4L}](1,1,-1,-1)\,,\hspace{0.5cm} \nu_{4L}(1,1,1,-1)\,,\nonumber\\
 & H(1,2,1,0)\,,\hspace{0.5cm} \phi(1,1,-2,2)\,,\hspace{0.5cm} S^{-}(1,1,-2,0)\,,\hspace{0.5cm} \Delta(1,\overline{3},0,2)\,.
\end{alignat}
 All kinds of the notations are self-explanatory as usual. $X_{\mu}$ and $Y_{\mu}$ are two new vector gauge fields related to the gauge subgroups $U(1)_{X}$ and $U(1)_{B-L}$, respectively. The last two numbers in the brackets are exactly the $U(1)_{X}$ and $U(1)_{B-L}$ charges. The fermion subscript $i$ indicates the first three generation fermions, in addition, the model newly includes the fourth generation quarks and leptons. It should also be noted that the fourth generation left-handed fermions are all singlets under $SU(2)_{L}$. The right-handed neutrinos are absent in (1). The reason for this is that the neutral fermions $\nu_{iL},\nu_{4L}$ are considered as Majorana particles, so the corresponding $\nu_{R}$ are not independent fields but rather determined by the relation $\nu_{R}=C\,\overline{\nu_{L}}^{T}$, in which $C$ is a charge conjugation matrix. Besides the SM doublet Higgs field, two complex singlet scalar $\phi$ and $S^{-}$, and a symmetric triplet scalar $\Delta$ are introduced in the model. After the gauge symmetry breakings, $\phi$ and $S^{-}$ have respectively a zero electric charge and a negative electric charge. These scalar field representations are such as
\ba
 \phi=\frac{\phi^{0}+i\phi'^{0}}{\sqrt{2}}\,,\hspace{0.5cm}
 H=\left(\begin{array}{c}H^{+}\\H^{0}\end{array}\right),\hspace{0.5cm}
 \Delta=\left(\begin{array}{cc}\Delta^{0}&\frac{\Delta^{+}}{\sqrt{2}}\\\frac{\Delta^{+}}{\sqrt{2}}&\Delta^{++}\end{array}\right).
\ea
 In addition, $\widetilde{H}=i\tau_{2}H^{*}$ will also be referred. In brief, all the non-SM particles in (1) compose the new particle spectrum, and they play key roles in the new physics beyond the SM. In general, the new particle masses are the TeV scale or above it, so they should appear in the TeV scale circumstances, for example, in the early universe.

 Under the model gauge symmetries, the invariant Lagrangian of the model is composed of the three parts in the following. Firstly, the gauge kinetic energy terms are
\begin{alignat}{1}
 \mathscr{L}_{Gauge}=
 &\:\mathscr{L}_{pure\,gauge}+i\overline{f}\gamma^{\mu}D_{\mu}f+(D^{\mu}H)^{\dagger}(D_{\mu}H)\nonumber\\
 & +(D^{\mu}\phi)^{\dagger}(D_{\mu}\phi)+(D^{\mu}S^{-})^{\dagger}(D_{\mu}S^{-})
   +Tr[(D^{\mu}\Delta)^{\dagger}(D_{\mu}\Delta)]\,,
\end{alignat}
 where $f$ stands for all kinds of the fermions in (1). The covariant derivative $D_{\mu}$ is defined as
\ba
 D_{\mu}=\partial_{\mu}+i\left(g_{3}G_{\mu}^{a}\frac{\lambda^{a}}{2}+g_{2}W_{\mu}^{i}\frac{\tau^{i}}{2}
 +g_{X}X_{\mu}\frac{Q_{X}}{2}+g_{B-L}Y_{\mu}\frac{B-L}{2}\right),
\ea
 where $\lambda^{a}$ and $\tau^{i}$ are Gell-Mann and Pauli matrices, $Q_{X}$ and $B-L$ are respectively the charge operators of $U(1)_{X}$ and $U(1)_{B-L}$, and $g_{3},g_{2},g_{X},g_{B-L}$ are four gauge coupling constants. After the gauge symmetry breakings, some gauge fields will generate their masses and mixings by the Higgs mechanism.

 Secondly, the model scalar potential is given by
\begin{alignat}{1}
  V_{Scalar}=
 &\:\mu_{H}^{2}H^{\dagger}H+\lambda_{H}(H^{\dagger}H)^{2}+\mu_{\phi}^{2}\phi^{\dagger}\phi+\lambda_{\phi}(\phi^{\dagger}\phi)^{2}\nonumber\\
 &+M_{S}^{2}S^{+}S^{-}+\lambda_{S}(S^{+}S^{-})^{2}+M_{\Delta}^{2}Tr[\Delta^{\dagger}\Delta]+\lambda_{\Delta}(Tr[\Delta^{\dagger}\Delta])^{2}\nonumber\\
 &+\lambda_{1}(H^{\dagger}H)(\phi^{\dagger}\phi)+\left(\lambda_{2}(H^{\dagger}H)+\lambda_{3}(\phi^{\dagger}\phi)\right)Tr[\Delta^{\dagger}\Delta]\nonumber\\
 &+\left(\lambda_{4}(H^{\dagger}H)+\lambda_{5}(\phi^{\dagger}\phi)+\lambda_{6}Tr[\Delta^{\dagger}\Delta]\right)(S^{+}S^{-})\nonumber\\
 &+\lambda_{7}(\widetilde{H}^{T}\Delta\widetilde{H}\phi^{\dagger}+h.c.)\,,
\end{alignat}
 where all kinds of the coupling parameters are self-explanatory. In general, the size of the coupling coefficients are $[\lambda_{1}, \lambda_{2}, \ldots, \lambda_{7}]\sim0.1$. For $[M_{S}^{2},M_{\Delta}^{2}]>0$, $M_{S}$ and $M_{\Delta}$ are original masses of the particles $S^{\pm}$ and $\Delta$, respectively. Their values are $M_{S}\sim 1$ TeV and $M_{\Delta}\sim 10^{5}$ TeV in the model. For $[\mu_{H}^{2},\mu_{\phi}^{2}]<0$, both the singlet $\phi$ and the doublet $H$ will develop non-zero VEVs. $\phi$ is responsible for the $B-L$ breaking $U(1)_{X}\otimes U(1)_{B-L}\rightarrow U(1)_{Y}$, in succession, $H$ completes the electroweak breaking $SU(2)_{L}\otimes U(1)_{Y}\rightarrow U(1)_{em}$. The triplet $\Delta$ also takes part in the two breakings because it is involved in the couplings of the last term in (5). However, it is about to be seen that $\Delta$ has only a very tiny VEV due to the heavy $M_{\Delta}$, therefore $\Delta$ only plays a secondary role in these breakings but it plays a key role in the generation of the tiny neutrino masses. Lastly, the singlet $S^{-}$ does not participate in any breakings because of its VEV vanishing. In comparison with the SM Higgs sector \cite{23}, the phenomena of the model scalar sector are variety and more interesting.

 Thirdly, the model Yukawa couplings are written as
\begin{alignat}{1}
  \mathscr{L}_{Yukawa}=
 & \left(\overline{Q_{i}}\widetilde{H},\overline{u_{4L}}M_{u_{4}}\right)
   \left(\begin{array}{cc}y^{u}_{1}I+y^{u}_{2}Y^{u}&y^{u}_{3}F^{u}\\y^{u}_{3}F^{u\dagger}&-1\end{array}\right)
   \left(\begin{array}{c}u_{iR}\\u_{4R}\end{array}\right)\nonumber\\
 &+\left(\overline{Q_{i}}H,\overline{d_{4L}}M_{d_{4}}\right)
   \left(\begin{array}{cc}y^{d}_{1}I+y^{d}_{2}Y^{d}&y^{d}_{3}F^{d}\\y^{d}_{3}F^{d\dagger}&-1\end{array}\right)
   \left(\begin{array}{c}d_{iR}\\d_{4R}\end{array}\right) \nonumber\\
 &+\left(\overline{L_{i}}H,\overline{e_{4L}}M_{e_{4}}\right)
   \left(\begin{array}{cc}y^{e}_{1}I+y^{e}_{2}Y^{e}&y^{e}_{3}F^{e}\\y^{e}_{3}F^{e\dagger}&-1\end{array}\right)
   \left(\begin{array}{c}e_{iR}\\e_{4R}\end{array}\right) \nonumber\\
 &+\frac{1}{2}\left(\overline{L_{i}}\Delta,\overline{\nu_{4L}}\phi\right)
   \left(\begin{array}{cc}y^{\nu}_{1}I+y^{\nu}_{2}Y^{\nu}&0\\0&y_{\nu_{4}}\end{array}\right)
   \left(\begin{array}{c}C\overline{L_{i}}^{T}\\C\overline{\nu_{4L}}^{T}\end{array}\right) \nonumber\\
 &+y^{u}_{4}\overline{d_{4L}}S^{-}O^{u\dagger}u_{R}+y^{d}_{4}\overline{u_{4L}}S^{+}O^{d\dagger}d_{R}
  +y^{e}_{4}\overline{\nu_{4L}}S^{+}O^{e\dagger}e_{R}+h.c.\,.
\end{alignat}
 These Yukawa couplings have apparently uniform and regular frameworks. $M_{f_{4}}$ are exactly the masses of the fourth generation quarks and charged lepton, which are directly allowed by the model symmetries, however, $M_{f_{4}}$ are about few TeV based on the model theoretical consistency. The coupling coefficients, $y^{f}_{1},y^{f}_{2},y^{f}_{3},y^{f}_{4}$, are chosen some real numbers since their complex phases can be absorbed by the redefined fermion fields. Since every coupling coefficients scale the order of magnitude of itself term, they are arranged such a hierarchy as $|y^{f}_{4}|<|y^{f}_{1}|<|y^{f}_{2}|<|y^{f}_{3}|\leqslant1$ in the model. The coupling matrices, $I,Y^{f},F^{f},O^{f}$, characterize the flavor mixings among the four generation fermions. At present we are lack of an underlying understand for the fermion flavor origin, however, it is believed that some flavor family symmetry is embedded in an underlying theory at a certain high-energy scale, but it is broken at the low-energy scale. The coupling matrices should imply some information of the flavor symmetry and its breaking. Therefore I suggest that the flavor structures of the coupling matrices have such a style as
\begin{alignat}{1}
 & I=\left(\begin{array}{ccc}1&0&0\\0&1&0\\0&0&1\end{array}\right),\hspace{0.5cm}
   Y^{f=u,d,e,\nu}=\left(\begin{array}{ccc}0&a_{f}&-a_{f}\\a_{f}&1&1\\-a_{f}&1&1\end{array}\right),\nonumber\\
 & F^{f=u,d,e}=\left(\begin{array}{c}0\\b_{f}\\1\end{array}\right),\hspace{0.5cm}
   O^{f=u,d,e}=\left(\begin{array}{c}0\\c_{1f}\\c_{2f}\\1\end{array}\right).
\end{alignat}
 The flavor structures are both simple and reasonable, in particular, there are only few flavor parameters. The size of the flavor parameters are normally $[|a_{f}|,|b_{f}|,|c_{1f}|,|c_{2f}|]$ $\sim 0.1$. The majority of their complex phases can be removed by the redefined fermion fields. The remaining complex phases will become the sources of the $CP$ violations in the quark and lepton sectors. In (6), the $y^{f}_{1}I$ couplings have evidently a full flavor symmetry among the first three generation fermions but they are relatively smaller. The $y^{f}_{2}Y^{f}$ couplings only keep such a discrete symmetry $S_{2}$ as $f_{2}\rightleftharpoons-f_{3}$ between the second and third generation fermions. The couplings between the first three and fourth generation fermions, $y^{f}_{3}F^{f}$, break the flavor symmetry $S_{2}$, but they are relatively bigger. Lastly, the couplings involving the charged scalar $S^{\pm}$, $y^{f}_{4}O^{f}$, are the smallest ones. After the gauge symmetries are broken spontaneously, all kinds of the fermion masses, $m_{u,d.e}$, $m_{\nu}$, $M_{\nu_{4}}$, will be generated by the corresponding couplings and the VEVs of $H,\,\Delta,\,\phi$, respectively. Finally, I in particular point out that the matrices $y^{f}_{1}I+y^{f}_{2}Y^{f}$ can be diagonalized by the unitary matrix $U_{0}$ as follows
\begin{alignat}{1}
 & U_{0}^{T}(y^{f}_{1}I+y^{f}_{2}Y^{f})U_{0}=
  \left(\begin{array}{ccc}y^{f}_{1}-\sqrt{2}a_{f}y^{f}_{2}&0&0\\0&y^{f}_{1}+\sqrt{2}a_{f}y^{f}_{2}&0\\0&0&y^{f}_{1}+2y^{f}_{2}\end{array}\right),\nonumber\\
 & U_{0}=\frac{1}{2}\left(\begin{array}{ccc}\sqrt{2}&\sqrt{2}&0\\-1&1&\sqrt{2}\\1&-1&\sqrt{2}\end{array}\right).
\end{alignat}
 The mixing angles of $U_{0}$ are $\theta_{12}=\theta_{23}=45^{\circ},\theta_{13}=0^{\circ}$. It evidently distinguishes from the tri-bimaximal mixing matrix \cite{24}. For $|y^{f}_{1}|<|a_{f}y^{f}_{2}|<|y^{f}_{2}|$, the first and second eigenvalues are approximately the same size, and the third one is the biggest. This property of (8) plays a key role in the neutrino mass and mixing in the model.

 In summary, the above features of the particle contents and Lagrangian are very important not only for the particle masses and mixings, but also guarantee the cold dark matter and leptogenesis in the model. In the following sections of the paper, it is about to be seen that $\nu_{4L}$ has unique natures and plays a special role in the model. It is actually a cold dark matter particle. The leptogenesis is really implemented by the decay $S^{-}\rightarrow e^{-}_{i}+\overline{\nu_{4}}$. In a word, the above contents form the theoretical framework of the model.

\vspace{1cm}
 \noindent\textbf{III. Symmetry Breakings and Particle Masses and Mixings}

\vspace{0.3cm}
 The gauge symmetry breakings of the model go through two stages. The first step of the breakings is $U(1)_{X}\otimes U(1)_{B-L}\rightarrow U(1)_{Y}$, namely the $B-L$ breaking, in succession, the second step is $SU(2)_{L}\otimes U(1)_{Y}\rightarrow U(1)_{em}$, i.e. the electroweak breaking. The former is achieved by the real part component of $\phi$ developing a non-vanishing VEV at the TeV scale, while the latter is accomplished by the neutral component of $H$ developing a non-vanishing VEV at the electroweak scale. In addition, the neutral component of $\Delta$ also develops a very tiny but non-zero VEV owing of the last term couplings in (5). The scalar vacuum structures and the conditions of the vacuum stabilization are easy derived from the scalar potential (5). The details are as follows
\begin{alignat}{1}
 & \phi \rightarrow \frac{\phi^{0}+v_{\phi}}{\sqrt{2}}\,,\hspace{0.3cm}
  H \rightarrow \left(\begin{array}{c}0\\\frac{H^{0}+\,v_{H}}{\sqrt{2}}\end{array}\right),\hspace{0.3cm}
  \Delta\rightarrow\left(\begin{array}{cc}\frac{\Delta^{0}+v_{\Delta}}{\sqrt{2}}&\frac{\Delta^{+}}{\sqrt{2}}\\\frac{\Delta^{+}}{\sqrt{2}}&\Delta^{++}\end{array}\right),\hspace{0.3cm}
  S^{-} \rightarrow S^{-},\nonumber\\
 & \langle\phi\rangle=\frac{v_{\phi}}{\sqrt{2}}
   =\sqrt{\frac{\lambda_{1}\mu_{H}^{2}-2\lambda_{H}\mu_{\phi}^{2}}{4\lambda_{\phi}\lambda_{H}-\lambda_{1}^{2}}}\,,\hspace{0.5cm} \langle H\rangle=\frac{v_{H}}{\sqrt{2}}
   =\sqrt{\frac{\lambda_{1}\mu_{\phi}^{2}-2\lambda_{\phi}\mu_{H}^{2}}{4\lambda_{\phi}\lambda_{H}-\lambda_{1}^{2}}}\,,\nonumber\\
 &\langle\Delta\rangle=\frac{v_{\Delta}}{\sqrt{2}}=\frac{-\lambda_{7}v_{\phi}v_{H}^{2}}{\sqrt{2}(2M_{\Delta}^{2}+\lambda_{2}v_{H}^{2}+\lambda_{3}v_{\phi}^{2})}
   \approx\frac{-\lambda_{7}v_{\phi}v_{H}^{2}}{2\sqrt{2}M_{\Delta eff}^{2}}\,,\hspace{0.5cm} \langle S^{-}\rangle=0\,,
\end{alignat}
 The stable conditions include $[\mu_{H}^{2},\mu_{\phi}^{2},\lambda_{1}]<0$, $[\lambda_{H},\lambda_{\phi},\lambda_{S},\lambda_{\Delta},\lambda_{2},\lambda_{3},\lambda_{4},\lambda_{5},\lambda_{6},4\lambda_{\phi}\lambda_{H}-\lambda_{1}^{2}]>0$, and $M_{S}^{2}\gg [v_{\phi}^{2},v_{H}^{2}]$. In addition, $|\lambda_{1}|$ should be sufficient small so that $v_{\phi}$ is one order of magnitude larger than $v_{H}$, in this way, this ensures that the $B-L$ breaking precedes the electroweak breaking. $M_{\Delta eff}$ in (9) is a effective mass of the $\Delta$ particle when the breakings are completed (see (10)). Provided that $M_{\Delta}\sim10^{5}$ TeV, $v_{\phi}\sim2.5$ TeV, $v_{H}\sim250$ GeV, and $\lambda_{7}\sim0.1$, thus this naturally leads to $v_{\Delta}\sim 0.1$ eV, consequently, gives the tiny neutrino masses. Thus it can be seen that the tiny nature of the neutrino masses essentially originates in the very heavy $M_{\Delta}$ in the model. In this sense, this is a new form of the seesaw mechanism \cite{25}. Finally, the $S^{-}$ field has a vanishing VEV, so it does not actually participate in the breakings. In short, all the conditions are not difficult to be satisfied so long as the parameters are chosen as some suitable values.

 After the model gauge symmetry breakings are over, the following massive scalar bosons, $H^{0}$, $\phi^{0}$, $S^{\pm}$, $\Delta[\Delta^{0},\Delta^{\pm},\Delta^{\pm\pm}]$, appear in the scalar sector. Their masses and mixings are such as
\begin{alignat}{1}
 & tan2\theta_{h}=\frac{\lambda_{1}v_{\phi}v_{H}}{\lambda_{\phi}v_{\phi}^{2}-\lambda_{H}v_{H}^{2}}\,,\nonumber\\
 & M_{H^{0},\phi^{0}}^{2}=\left(\lambda_{\phi}v_{\phi}^{2}+\lambda_{H}v_{H}^{2}\right)
  \mp\left|\lambda_{\phi}v_{\phi}^{2}-\lambda_{H}v_{H}^{2}\right|\sqrt{1+tan^{2}2\theta_{h}}\,,\nonumber\\
 & M_{S eff}^{2}=M_{S}^{2}+\frac{1}{2}(\lambda_{4}v_{H}^{2}+\lambda_{5}v_{\phi}^{2}+\lambda_{6}v_{\Delta}^{2}),\nonumber\\
 & M_{\Delta eff}^{2}=M_{\Delta}^{2}+\frac{1}{2}(\lambda_{2}v_{H}^{2}+\lambda_{3}v_{\phi}^{2}+2\lambda_{\Delta}v_{\Delta}^{2}),
\end{alignat}
 where $\theta_{h}$ is the mixing angle between $H^{0}$ and $\phi^{0}$. Provided that $\lambda_{\phi}\sim\lambda_{H}>|\lambda_{1}|$, then $tan2\theta_{h}<0.1$, thus the two neutral boson masses are approximately $M_{H^{0}}\approx\sqrt{2\lambda_{H}}\,v_{H}$ and $M_{\phi^{0}}\approx\sqrt{2\lambda_{\phi}}\,v_{\phi}$. In a similar way, there is also a very weak mixing between $H^{0}$ and $\Delta^{0}$, or between $\phi^{0}$ and $\Delta^{0}$. However, these very weak mixings in the scalar sector can all be ignored throughout. $M_{S eff}$ and $M_{\Delta eff}$ are respectively the effective masses of the particles $S^{\pm}$ and $\Delta$. Provided that $M_{S}\leqslant v_{\phi}$, then $M_{S eff}\sim v_{\phi}$. In view of $M_{\Delta}\gg v_{\phi}$, obviously, there is $M_{\Delta eff}\approx M_{\Delta}$. At present, $M_{H^{0}}$ has been measured by the LHC \cite{26}, it's value is $125$ GeV. However, the model predicts that $M_{\phi^{0}}$ and $M_{S}$ are about $1$ TeV, it are quite feasible to find the two bosons at the LHC, but the $\Delta$ particles are too heavy to be detected.

 In the gauge sector, the gauge symmetry breakings give rise to masses and mixings for some of the vector gauge bosons through the Higgs mechanism. The breaking procedure of $U(1)_{X}\otimes U(1)_{B-L}\rightarrow U(1)_{Y}$ is such as
\begin{alignat}{1}
 & g_{X}X_{\mu}\frac{Q_{X}}{2}+g_{B-L}Y_{\mu}\frac{B-L}{2}\longrightarrow
   g_{1}(B_{\mu}\frac{Q_{Y}}{2}+Z'_{\mu}\frac{Q_{N}}{2}),\nonumber\\
 & g_{1}=\frac{g_{X}g_{B-L}}{\sqrt{g_{X}^{2}+g_{B-L}^{2}}}\,,\hspace{0.5cm} tan\theta_{g}=\frac{g_{X}}{g_{B-L}}\,,\nonumber\\
 & Q_{Y}=Q_{X}+(B-L)\,,\hspace{0.5cm} Q_{N}=-tan\theta_{g}Q_{X}+cot\theta_{g}(B-L)\,,\nonumber\\
 & B_{\mu}=cos\theta_{g}X_{\mu}+sin\theta_{g}Y_{\mu}\,,\hspace{0.5cm}
   Z'_{\mu}=-sin\theta_{g}X_{\mu}+cos\theta_{g}Y_{\mu}\,,\nonumber\\
 & M_{B_{\mu}}=0\,,\hspace{0.5cm} M_{Z'_{\mu}}=\frac{1}{2}|Q_{N}(\phi)|g_{1}v_{\phi}=\frac{2g_{1}v_{\phi}}{sin2\theta_{g}}\,,
\end{alignat}
 where $g_{1},B_{\mu},Q_{Y}$ are respectively the gauge coupling constant, gauge field, supercharge operator of $U(1)_{Y}$, $Z'_{\mu}$ is an obtained mass neutral gauge field, and $Q_{N}$ is a new charge operator related to $Z'_{\mu}$. There are two gauge parameters $g_{1}$ and $tan\theta_{g}$ in (11), however, $g_{1}$ is not a free parameter but determined by the electroweak relation $g_{1}=\sqrt{4\pi\alpha}/cos\theta_{W}$, only the mixing angle $tan\theta_{g}$ is a free parameter. In addition, the last equation in (11) implies $M_{Z'_{\mu}}\geqslant2g_{1}v_{\phi}$, so $M_{Z'_{\mu}}$ should be few TeV or so. It should also be pointed out that the mixing between $Z'_{\mu}$ and $Z_{\mu}$, which is the SM weak neutral gauge boson, is very small. Their mixing angle is given by
\ba
 tan2\theta'_{g}\approx\frac{sin^{3}\theta_{g}cos\theta_{g}v_{H}^{2}}{2sin\theta_{W}v_{\phi}^{2}}\,.
\ea
 Because of $v_{H}^{2}/v_{\phi}^{2}\sim10^{-2}$ and $sin\theta_{W}\sim0.5$, this mixing can indeed be ignored.

 Below the electroweak scale, the Yukawa sector becomes clear and simple since the fourth generation heavy quarks and charged lepton have decoupled. After they are integrated out from (6), the effective Yukawa coupling matrices of the three generation quarks and charged lepton are given by
\ba
 Y^{f=u,d,e}_{eff}=y^{f}_{1}I+y^{f}_{2}Y^{f}+(y^{f}_{3})^{2}F^{f}\otimes F^{f\dagger},\hspace{0.5cm}
 F^{f}\otimes F^{f\dagger}=\left(\begin{array}{ccc}0&0&0\\0&b_{f}b_{f}^{*}&b_{f}\\0&b_{f}^{*}&1\end{array}\right).
\ea
 Each term physical meaning is very explicit. According to the standard procedures, in (6) the symmetry breakings give rise to all kinds of the fermion masses as follows
\ba
  M_{f=u,d,e}=-\frac{v_{H}}{\sqrt{2}}Y^{f}_{eff}\,,\hspace{0.5cm}
  M_{\nu}=-\frac{v_{\Delta}}{\sqrt{2}}(y^{\nu}_{1}I+y^{\nu}_{2}Y^{\nu}),\hspace{0.5cm}
  M_{\nu_{4}}=-\frac{v_{\phi}}{\sqrt{2}}\,y_{\nu_{4}}\,.
\ea
 Obviously, there is the mass hierarchical relation $M_{\nu}\sim 0.01\,\mathrm{eV}\ll M_{f=u,d,e}<M_{\nu_{4}}\sim 1\,\mathrm{TeV}$. In addition, the hierarchical coefficients, $|y^{f}_{1}|\ll |y^{f}_{2}|\ll |y^{f}_{3}|$, will lead to the hierarchical masses of the three generation quarks and charged lepton. On the other hand, there is not such a term as $F^{\nu}\otimes F^{\nu\dagger}$ in $M_{\nu}$, so $M_{\nu}$ is distinguished from $M_{f=u,d,e}$\,. This is a primary source that the lepton flavor mixing is greatly different from the quark one. In short,
 the interesting features of the fermion mass matrices dominate the fermion masses and flavor mixings.

 Finally, the fermion mass eigenvalues and flavor mixing matrices are solved by diagonalizing the above mass matrices. The quark and charged lepton mass matrices $M_{f=u,d,e}$ are hermitian, while the light neutrino mass matrix $M_{\nu}$ is symmetry. Therefore, the mass matrix diagonalizations are accomplished as such
\begin{alignat}{1}
 & U_{u}^{\dagger}M_{u}U_{u}=\mathrm{diag}\left(m_{u},m_{c},m_{t}\right),\hspace{0.5cm}
   U_{d}^{\dagger}M_{d}U_{d}=\mathrm{diag}\left(m_{d},m_{s},m_{b}\right),\nonumber\\
 & U_{e}^{\dagger}M_{e}U_{e}=\mathrm{diag}\left(m_{e},m_{\mu},m_{\tau}\right),\hspace{0.5cm}
   U_{n}^{\dagger}M_{\nu}U_{n}^{*}=\mathrm{diag}\left(m_{n_{1}},m_{n_{2}},m_{n_{3}}\right).
\end{alignat}
 In the light of the characteristic structures of $M_{f=u,d,e}$, the mass eigenvalues of the quarks and charged lepton are certainly some hierarchy, and the flavor mixing matrices $U_{f=u,d,e}$ are all closed to an unit matrix. In contrast, an exact solution of the $M_{\nu}$ diagonalization can be given by use of (8) as
\ba
 U_{n}=U_{0}\,,\hspace{0.5cm} m^{2}_{n_{2}}-m^{2}_{n_{1}}=2\sqrt{2}y^{\nu}_{1}y^{\nu}_{2}a_{\nu}v_{\Delta}^{2}\,,\hspace{0.5cm}
 m^{2}_{n_{3}}-m^{2}_{n_{2}}\approx 2(y^{\nu}_{2}v_{\Delta})^{2}.
\ea
 Obviously, $U_{n}$ is completely different from $U_{f=u,d,e}$\,, moreover, the two mass-squared differences can explain the neutrino data very well. The above results are convenient for the following numerical analysis. The flavor mixing matrix in the quark sector and one in the lepton sector are respectively defined by \cite{27,28}
\ba
 U_{u}^{\dagger}\,U_{d}=U_{CKM}\,,\hspace{0.5cm}
 U_{e}^{\dagger}\,U_{n}=U_{PMNS}\cdot\mathrm{diag}\left(e^{i\beta_{1}},e^{i\beta_{2}},1\right)\,.
\ea
 The two unitary matrices $U_{CKM}$ and $U_{PMNS}$ are parameterized by the standard form in particle data group \cite{1}. $\beta_{1}, \beta_{2}$ are two Majorana phases in the lepton mixing. All kinds of the mixing angles and $CP$-violating phases can be calculated numerically. Finally, all of the results can be compared with the current and future experimental data.

\vspace{1cm}
 \noindent\textbf{IV. Cold Dark Matter and Leptogenesis}

\vspace{0.3cm}
 The model can naturally and elegantly account for the cold dark matter and leptogenesis after the model symmetry breakings are completed. The fourth generation Majorana neutrino $\nu_{4L}$ own some unique properties. It has been seen from the model lagrangian that $\nu_{4L}$ has only the three types of the couplings, $\overline{\nu_{4L}}\gamma^{\mu}\nu_{4L}Z'_{\mu}$, $\nu_{4L}^{T}\nu_{4L}\phi^{0}$, and $\overline{e_{R}}\nu_{4L}S^{-}$. Provided that the mass order $M_{\nu_{4}}<M_{S}<[M_{u_{4}},M_{d_{4}},M_{e_{4}}]$, the only decay channel of $\nu_{4L}$ is $\nu_{4L}\rightarrow e_{i}u_{j}\overline{d_{k}}$ via an off-shell boson $S^{+}$ because there are some weak mixings between the fourth generation quarks and the first three generation ones. If the coupling coefficients are $|y^{f}_{4}|\leqslant10^{-9}$ in the last line of (6), then the $\nu_{4L}$ decay width is estimated as $\leqslant10^{-44}$ GeV. In other words, the $\nu_{4L}$ lifetime is actually two orders of magnitude longer than the now age of universe, therefore it becomes a relatively stable particle in the universe. On the other hand, a pair of $\nu_{4L}$ can annihilate into other particle pair. The annihilate processes are mediated by either the gauge boson $Z'_{\mu}$ or the scalar boson $\phi^{0}$. Because the $\nu_{4L}$ mass is derived from the $B-L$ breaking, $M_{\nu_{4}}$ should be around one TeV. Consequently, the Majorana neutrino $\nu_{4L}$ is genuinely a weak interactive massive particle (WIMP), of course, it also belongs to one of the fewer species of particles which can survive from the early universe to the now epoch. Therefore $\nu_{4L}$ is a good candidate of the cold dark matter \cite{29}.

 The annihilate channels of $\nu_{4L}$ have two ways. The principal annihilate process is that a pair of $\nu_{4L}$ annihilate into all kinds of the SM fermion pairs by the intermediate gauge boson $Z'_{\mu}$, as shown in the figure (1).
\begin{figure}
 \centering
 \includegraphics[totalheight=4cm]{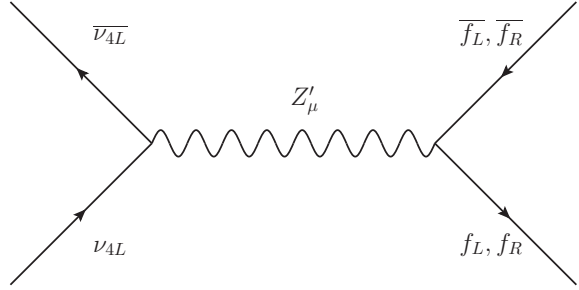}
 \caption{The diagram of the cold dark matter $\nu_{4L}$ annihilating into all kinds of the SM fermion pairs by the gauge boson $Z'_{\mu}$ at the TeV scale\,.}
\end{figure}
 The other annihilate process is that two $\nu_{4L}$ neutrinos annihilate into two Higgs bosons $H^{0}$ by the TeV scale boson $\phi^{0}$ mediating. Because $\lambda_{1}$ is smaller, the cross section of the latter case is normally far smaller than one of the former except for some Breit-Wigner resonance points, so I can ignore the latter and only consider the former. The annihilate cross section of the figure (1) is calculated as follows
\begin{alignat}{1}
 & \sigma(\nu_{4L}+\overline{\nu_{4L}}\rightarrow f+\overline{f})=\frac{g_{1}^{4}\,Q_{N}^{2}(\nu_{4L})\,s}
   {256\,\pi[(s-M_{Z'_{\mu}}^{2})^{2}+(\Gamma_{Z'_{\mu}}M_{Z'_{\mu}})^{2}]}
   \sum\limits_{f_{L},f_{R}}\sqrt{\frac{s-4m_{f}^{2}}{s-4M_{\nu_{4}}^{2}}}R_{f}\,,\nonumber\\
 &R_{f}=\frac{Q_{N}^{2}(f_{L})+Q_{N}^{2}(f_{R})}{3}(1-\frac{m_{f}^{2}+M_{\nu_{4}}^{2}}{s}+\frac{4m_{f}^{2}M_{\nu_{4}}^{2}}{s^{2}})\nonumber\\
 & +2Q_{N}(f_{L})Q_{N}(f_{R})\frac{m_{f}^{2}}{s}(1-\frac{2M_{\nu_{4}}^{2}}{s})
   +\left(Q_{N}(f_{L})-Q_{N}(f_{R})\right)^{2}\frac{m_{f}^{2}M_{\nu_{4}}^{2}}{M_{Z'_{\mu}}^{4}}(1-\frac{2M_{Z'_{\mu}}^{2}}{s}),\nonumber\\
 & \Gamma(Z'_{\mu}\rightarrow f+\overline{f})=\frac{g_{1}^{2}M_{Z'_{\mu}}}{96\pi}\sum\limits_{f_{L},f_{R},\nu_{4L}}
   \sqrt{1-\frac{4m_{f}^{2}}{M_{Z'_{\mu}}^{2}}}R'_{f}\,,\nonumber\\
 & R'_{f}=\left(Q_{N}^{2}(f_{L})+Q_{N}^{2}(f_{R})\right)(1-\frac{m_{f}^{2}}{M_{Z'_{\mu}}^{2}})
   +6Q_{N}(f_{L})Q_{N}(f_{R})\frac{m_{f}^{2}}{M_{Z'_{\mu}}^{2}}\,,
\end{alignat}
 where $s=4M_{\nu_{4}}^{2}/(1-v^{2})$ is the squared center-of-mass energy, $v$ is the velocity of $\nu_{4L}$ in the center-of-mass frame. The sum for $f_{L},f_{R}$ count all kinds of the SM fermions who are permitted by kinematics. In fact, all of the relatively lighter $m_{f}$ in (18) can been approximated to zero except for the relatively heavier $m_{t}, M_{\nu_{4}}$. On the basis of WIMP, the relic abundance of $\nu_{4L}$ in the current universe is determined by the annihilation cross section as such
\ba
 \Omega h^{2}\approx\frac{2.58\times 10^{-10}\,\mathrm{GeV^{-2}}}{\langle\sigma v_{r}\rangle}\,,
\ea
 where $v_{r}$ is the relative velocity of the two annihilate particles. In addition, the heat average (19) can be calculated by $\langle\sigma v_{r}\rangle\approx a+b\langle v^{2}\rangle=a+b\frac{3T_{f}}{M_{\nu_{4}}}$, in which $T_{f}\approx M_{\nu_{4}}/20$ is the freeze temperature of $\nu_{4L}$. A rough estimate is as follows. Because of $g_{1}^{2}\sim0.1,\, \sum Q^{2}_{N}(f)\sim10,\, \sqrt{ s}\sim M_{Z'_{\mu}}\sim 10^{3}\,\mathrm{GeV}$, a weak cross section is naturally obtained as $\sigma\sim10^{-9}\,\mathrm{GeV^{-2}}$, eventually, it leads to $\Omega h^{2}\sim0.1$, which is closed to the observation value.

 The baryon asymmetry through the leptogenesis can be implemented in the model. On the basis of the relevant couplings in (5) and (6), the main decay channel of the charged scalar boson $S^{-}$ is $S^{-}\rightarrow \overline{\nu_{4}}e^{-}_{i}$ (not including a heavy $e^{-}_{4}$), in addition, $S^{-}\rightarrow \overline{u_{i}}d_{j}$ can be ignored because its decay branching ratio is smaller. The figure (2) draws the tree and loop diagrams of $S^{-}\rightarrow \overline{\nu_{4}}e^{-}_{i}$, of course, its $CP$ conjugate process $S^{+}\rightarrow \nu_{4}\overline{e^{-}_{i}}$ has the corresponding diagrams.
\begin{figure}
 \centering
 \includegraphics[totalheight=4cm]{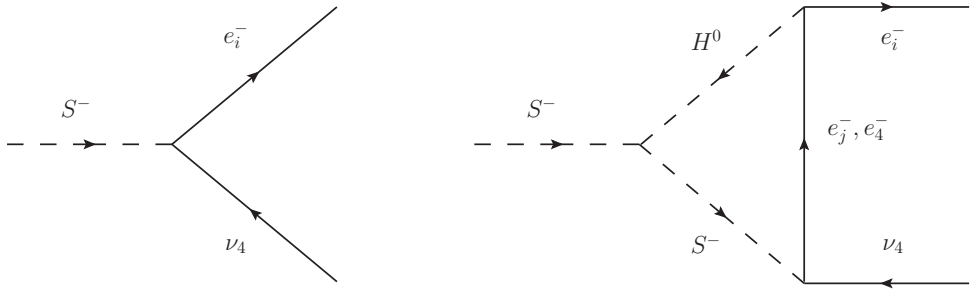}
 \caption{The tree and loop diagrams of the decay of $S^{-}\rightarrow e^{-}_{i}+\overline{\nu_{4}}$, by which the $B-L$ asymmetry is generated\,.}
\end{figure}
 However, this decay process has simultaneously three items of the notable characteristics. Firstly, the decay evidently violates one unit of the $B-L$ quantum number, namely $\bigtriangleup(B-L)=-1$. Secondly, because there is a $CP$-violating source in the leptonic Yukawa sector, a $CP$ asymmetry of the decay is surely generated by the interference between the tree diagram and the loop one. The $CP$ asymmetry is defined and calculated as
\begin{alignat}{1}
 \varepsilon & =\frac{\Gamma(S^{-}\rightarrow e^{-}_{i}+\overline{\nu_{4}})-\Gamma(S^{+}\rightarrow e^{+}_{i}+\nu_{4})}
                 {\Gamma(S^{-}\rightarrow e^{-}_{i}+\overline{\nu_{4}})+\Gamma(S^{+}\rightarrow e^{+}_{i}+\nu_{4})}\nonumber\\
 & =\frac{-v_{H}\lambda_{4}\sum\limits_{i=1}^{3}\sum\limits_{j=1}^{4}M_{e_{i}}Im(O^{e}_{i}T^{*}_{ij}O^{e*}_{j})
  (M_{S}^{2}-M_{\nu_{4}}^{2}-M_{e_{j}}^{2})f_{j}}
   {4\sqrt{2}\,\pi\sum\limits_{i=1}^{3}|O^{e}_{i}|^{2}(M_{S}^{2}-M_{\nu_{4}}^{2})^{2}}\,,\nonumber\\
 & T_{ij}=(y^{e}_{1}I+y^{e}_{2}Y^{e})_{ij}\,,\hspace{0.5cm} T_{i4}=y^{e}_{3}F^{e},\nonumber\\
 & f_{j}=ln\frac{x_{j}}{x_{j}-1}\,,\hspace{0.5cm} x_{j}=\frac{M_{S}^{2}}{M_{\nu_{4}}^{2}}(1-\frac{M_{e_{j}}^{2}}{M_{S}^{2}-M_{\nu_{4}}^{2}})+\frac{M_{H^{0}}^{2}}{M_{S}^{2}-M_{\nu_{4}}^{2}}\,,
\end{alignat}
 where provided with $M_{e_{4}}>M_{S}+M_{\nu_{4}}$. Thirdly, the decay is really an out-of-equilibrium process. Provided that the coupling coefficient $|y^{e}_{4}|\leqslant10^{-9}$ as before, then the decay rate is far smaller than the Hubble expansion rate of the universe, namely
\ba
 \frac{\Gamma(S^{-}\rightarrow e^{-}_{i}+\overline{\nu_{4}})}{H(T=M_{S})}=
 \frac{\frac{M_{S}}{16\pi}(1-\frac{M_{\nu_{4}}^{2}}{M_{S}^{2}})^{2}\sum\limits_{i=1}^{3}|y^{e}_{4}O^{e}_{i}|^{2}}
 {\frac{1.66\sqrt{g_{*}}M_{S}^{2}}{M_{pl}}}\ll1\,,
\ea
 where $m_{e_{i}}^{2}/M_{S}^{2}$ has been approximated to zero, $M_{pl}=1.22\times10^{19}$ GeV, and $g_{*}$ is the effective number of relativistic degrees of freedom. In brief, the decay process $S^{-}\rightarrow e^{-}_{i}+\overline{\nu_{4}}$ can indeed satisfy Sakharov's three conditions \cite{30}. Therefore, this mechanism can naturally generate an asymmetry of $B-L$ at the TeV scale. It can be seen from (20) that the $CP$ asymmetry $\varepsilon$ linearly depends on the three quantities $\lambda_{4},\,T_{ij},\,f_{j}$, in particular, the major contribution comes from $T_{i4}$, namely the process by the $e^{-}_{4}$ inner mediating. In addition, $\varepsilon$ has no relation with $y^{e}_{4}$ though the decay rate depends on $y^{e}_{4}$. In the model, there are $\lambda_{4}\sim0.1,\,T_{i4}\sim0.01,\,f_{j}\sim0.1$, thus $M_{S}\sim1$ TeV will give $\varepsilon\sim10^{-8}$.

 The above $B-L$ asymmetry occurs at the TeV scale. At that time the heavy $S^{\mp}$ and $\nu_{4}$ have been the non-relativistic particles, but the produced leptons are truly the relativistic states, moreover, their energy are normally more than $100$ GeV. Therefore the sphaleron electroweak transition can smoothly put into effect \cite{31}. Consequently, the $B-L$ asymmetry will eventually be converted into an asymmetry of the baryon number through the sphaleron process. According to the standard discussions, the baryon asymmetry is determined by
\ba
 \eta_{B}=7.04\,c_{sp}Y_{B-L}=7.04\,c_{sp}\left(\kappa\frac{(-1)\,\varepsilon}{g_{*}}\right),
\ea
 where $7.04$ is a ratio of the entropy density to the photon number density, $c_{sp}=28/79$ is a coefficient of the sphaleron conversion, $Y_{B-L}$ stands for the $B-L$ asymmetry, which is related to $\varepsilon$ by the expression in the parentheses. $\kappa$ is a dilution factor, it can actually be approximated to $\kappa\approx1$ on account of the very weak decay rate. At the TeV scale, only the SM particles are the relativistic statuses, whereas the non-SM particles are the non-relativistic statuses, so $g_{*}$ exactly equal to the effective number of degrees of freedom of the SM particles, namely $g_{*}=106.75$. In short, the baryon asymmetry can be calculated by the relations of (20) and (22). The model can achieve the leptogenesis at the TeV scale.

\vspace{1cm}
 \noindent\textbf{V. Numerical Results}

\vspace{0.3cm}
 In the section I present the model numerical results. The model involves a number of the new parameters besides the SM ones. In the light of the III and IV section discussions, the parameters involved in the numerical calculations are collected together in the following. The gauge sector has the gauge coupling $g_{1}$ and the mixing angle $tan\theta_{g}$. The scalar sector includes the three VEVs, $v_{\phi},\,v_{H},\,v_{\Delta}$, the two scalar boson masses $M_{H^{0}},\,M_{S}$, and the scalar coupling $\lambda_{4}$. The Yukawa sector has $M_{e_{4}}$, all kinds of the coupling coefficients, and the flavor parameters, see (6) and (7). Among which, the three parameters, $g_{1},\,v_{H},\,M_{H^{0}}$, in fact belong to the SM parameters. Their values have been determined in the electroweak scale physics and by the recent LHC experiments \cite{26}, namely $g_{1}=0.356,\,v_{H}=246\,\mathrm{GeV},\,M_{H^{0}}=125\,\mathrm{GeV}$. In addition, the three parameters, $v_{\phi},\,v_{\Delta},\,M_{e_{4}}$, also belong to the fundamental parameters in the model. Based on an overall consideration, the above six parameters are fixed throughout to the following values,
\begin{alignat}{1}
 & g_{1}=0.356, \hspace{0.5cm} v_{H}=246\:\mathrm{GeV},\hspace{0.5cm} M_{H^{0}}=125\:\mathrm{GeV},\nonumber\\
 & v_{\phi}=2.5\:\mathrm{TeV},\hspace{0.5cm} v_{\Delta}=0.1\:\mathrm{eV},\hspace{0.5cm} M_{e_{4}}=2.5\:\mathrm{TeV}.
\end{alignat}
 The rest of the model parameters have to be determined by fitting all kinds of the experimental data.

 In the light of (6), the model Yukawa sector consists of the relatively independent quark sector and lepton one, so I will respectively discuss them. In the quark sector, the ten parameters in the following are involved in fitting the quark masses and mixing. Their input values are such as
\begin{alignat}{1}
 & y^{u}_{1}=-1.1\times10^{-5},\hspace{0.3cm} y^{u}_{2}=-7.5\times10^{-3},\hspace{0.3cm}
   y^{u}_{3}=1\,,\hspace{0.3cm} a_{u}=-0.056\,,\hspace{0.3cm} b_{u}=0.018\,,\nonumber\\
 & y^{d}_{1}=-1.1\times10^{-5},\hspace{0.3cm} y^{d}_{2}=5.55\times10^{-4},\hspace{0.3cm}
   y^{d}_{3}=0.153\,,\hspace{0.3cm} a_{d}=0.168\,,\hspace{0.3cm} b_{d}=-0.039\,i\,.
\end{alignat}
 In the above the choices of the imaginary parts of $a_{u},\,a_{d},\,b_{u},\,b_{d}$ have a certain degree of freedoms. Here I only choose $b_{d}$ as a pure imaginary number since this style is simple and the best in the fits, of course, it is exactly the source of the $CP$ violation in the quark sector. In short, this set of the values in (24) are reasonable, and also consistent with the previous discussions. According to the relevant relations in (13)--(17), the quark masses and mixing are solved as follows
\begin{alignat}{1}
 & m_{u}=0.00234\,,\hspace{0.5cm} m_{c}=1.274\,,\hspace{0.5cm} m_{t}=173\,,\nonumber\\
 & m_{d}=0.00481\,,\hspace{0.5cm} m_{s}=0.0952\,,\hspace{0.5cm} m_{b}=4.18\,, \nonumber\\
 & s^{\,q}_{12}=0.2252\,,\hspace{0.3cm} s^{\,q}_{23}=0.0411\,,\hspace{0.3cm} s^{\,q}_{13}=0.00354\,,\hspace{0.3cm}
   \delta^{\,q}=0.377\,\pi=67.8^{\circ},\nonumber\\
 & J_{CP}^{\,q}=2.95\times10^{-5},
\end{alignat}
 where mass is in GeV unit, $s_{\alpha\beta}=sin\theta_{\alpha\beta}$. In addition, the Jarlskog invariant $J_{CP}^{\,q}$, which measures the magnitude of the $CP$ violation, is also figured out by using the quark mixing angles and complex phase. It is very clear that the numerical results in (25) accurately fit all the current experimental data of the quark masses, mixing, and $CP$ violation \cite{1}. However, there are several points worthy pointed out in the successful fits. Firstly, the hierarchical parameters, $y^{f}_{3},\,y^{f}_{2},\,y^{f}_{1}$, respectively dominate the 3rd, 2nd, 1st generation fermion masses, the parameters $a_{d}$ and $a_{u}$ impact on the mixing angles $sin\theta_{12}$ and $sin\theta_{13}$, the parameters $b_{d}$ and $b_{u}$ are in charge of $sin\theta_{23}$ and $\delta$. Secondly, both the hierarchy of $y^{f}_{3},\,y^{f}_{2},\,y^{f}_{1}$ and the flavor structures of (7) jointly lead that the quark transform matrices $U_{u}$ and $U_{d}$ in (15) are both closed to an unit matrix, as a result, the quark mixing matrix $U_{CKM}$ has eventually three small mixing angles. Thirdly, since the quark masses and mixing angles have been measured to a certain precision, the variable scope of the parameter space are very narrow. Fourthly, there is no any fine-tuning in the fits. Fifthly, there are the parameter relations $y^{d}_{1}=y^{u}_{1}$, $a_{d}=-3a_{u}$, $a_{d}-a_{u}=\lambda_{c}$ in (24), which $\lambda_{c}=0.224$ is Cabibbo mixing angle. This should not be by coincidence, maybe there is an underlying reason for them.

 In the lepton sector, although the case of the charged lepton is similar to one of the quarks, the neutrino case makes great differences due to its particularity. It is in fact seen from (8) and (16) that $U_{n}$ has been fixed to $U_{0}$, and the three parameters, $y^{\nu}_{1},\,y^{\nu}_{2},\,a_{\nu}$, determine the three mass eigenvalues of the neutrinos. Therefore, only $y^{\nu}_{1},\,y^{\nu}_{2}$ are used to fit the two mass-squared differences of the neutrinos and $a_{\nu}$ can be fixed freely. The relevant parameters are chosen as follows
\begin{alignat}{1}
 & y^{e}_{1}=3.51\times10^{-5},\hspace{0.3cm} y^{e}_{2}=5.7\times10^{-4},\hspace{0.3cm}
   y^{e}_{3}=0.0974\,,\hspace{0.3cm} a_{e}=0.24\,,\hspace{0.3cm} b_{e}=-0.1\,i\,,\nonumber\\
 & y^{\nu}_{1}=0.088\,,\hspace{0.5cm} y^{\nu}_{2}=0.304\,,\hspace{0.5cm} a_{\nu}=0.1\,,
\end{alignat}
 where the pure imaginary $b_{e}$ is chosen as the source of the $CP$ violation in the lepton sector. Inputting (26) into the relevant equations, the lepton masses and mixing are calculated as follows
\begin{alignat}{1}
 &m_{e}=0.511\:\mathrm{MeV},\hspace{0.5cm} m_{\mu}=105.7\:\mathrm{MeV},\hspace{0.5cm} m_{\tau}=1777\:\mathrm{MeV},\nonumber\\
 & m_{n_{1}}=0.0032\:\mathrm{eV},\hspace{0.5cm} m_{n_{2}}=0.0093\:\mathrm{eV},\hspace{0.5cm}
   m_{n_{3}}=0.049\:\mathrm{eV},\nonumber\\
 & s^{\,l}_{12}=0.557\,,\hspace{0.5cm} s^{\,l}_{23}=0.654\,,\hspace{0.5cm} s^{\,l}_{13}=0.153\,,\hspace{0.5cm}
   \delta^{\,l}=\pi,\nonumber\\
 & \beta_{1}=-0.063\,\pi,\hspace{0.5cm} \beta_{2}=-0.063\,\pi\,.
\end{alignat}
 For a convenient comparison with the experimental data, the above results are converted into the commonly interested quantities in neutrino physics. They are such as
\begin{alignat}{1}
 & \triangle m^{2}_{21}=7.57\times10^{-5}\:\mathrm{eV^{2}},\hspace{0.5cm}
   \triangle m^{2}_{32}=2.34\times10^{-3}\:\mathrm{eV^{2}},\nonumber\\
 & sin^{2}2\theta^{\,l}_{12}=0.86\,,\hspace{0.5cm} sin^{2}2\theta^{\,l}_{23}=0.98\,,\hspace{0.5cm}
   sin^{2}2\theta^{\,l}_{13}=0.092\,,\nonumber\\
 & m_{\beta\beta}=0.006\:\mathrm{eV},\hspace{0.5cm} J_{CP}^{\,l}=0\,,
 \end{alignat}
 where $\triangle m^{2}_{\alpha\beta}=m^{2}_{n_{\alpha}}-m^{2}_{n_{\beta}}$, and $m_{\beta\beta}$ is the effective Majorana mass for neutrinoless double beta decay. These results in (27) and (28) are excellently in agreement with the recent experimental data \cite{1,32}, in particular, the value of $sin^{2}2\theta^{\,l}_{13}$ is identical with the new results of the neutrino oscillation at Daya Bay \cite{33}. In addition, the valid input values in (26) are only seven, but the output values in (27) are twelve, the model really shows a stronger prediction power. Firstly, the lepton $CP$-violating phase angle is $\pi$, as a result, the Jarlskog invariant $J_{CP}^{\,l}$ is zero, i.e. the lepton $CP$ violation is vanishing. Of course the reason for this arises from $\theta_{13}=0$ in $U_{0}$. It should be emphasized that if $a_{e}$ has a imaginary part, there can be a non-vanishing $CP$ violation in the lepton sector. Secondly, the two Majorana phases are the same but smaller. Thirdly, $m_{\beta\beta}=0.006$ eV is not too small, it is therefore promising to detected in the near future. Anyway, all the predicted results are expected to be tested in further neutrino experiments.

 In the scheme of (26), $a_{e}$ completely determines $sin\theta_{12}$ and $sin\theta_{13}$, while $b_{e}$ has only a weak effect on $sin\theta_{23}$.
\begin{figure}
 \centering
 \includegraphics[totalheight=9cm]{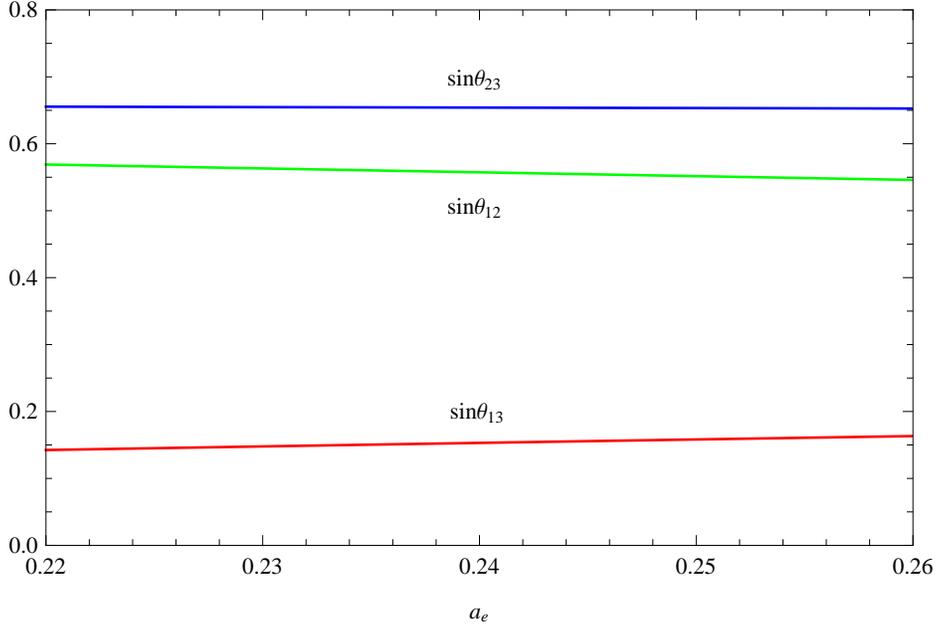}
 \caption{The graphs of the variations of the lepton mixing angles with the parameter $a_{e}$ when the other parameters are fixed in (26)\,.}
\end{figure}
 The figure (3) draws the variations of the three lepton mixing angles with the parameter $a_{e}$ when the other parameters are fixed in (26). It can be seen from the graphs that $sin\theta_{23}$ is almost unchanged, $sin\theta_{12}$ scales down very slowly as $a_{e}$ increasing, in the meantime, $sin\theta_{13}$ has only a weak rise toward right-hand side. In brief, the scenario can excellently explain that the lepton mixing angles are $sin\theta_{23}\approx41^{\circ}$, $sin\theta_{12}\approx34^{\circ}$ and $sin\theta_{13}\approx9^{\circ}$. All these results essentially stem from the theoretical structures of the model.

 I next calculate the relic abundance of the cold dark mater neutrino $\nu_{4L}$. The calculation of its annihilate cross section needs refer the two parameters $tan\theta_{g}$ and $y_{\nu_{4}}$. By virtue of the relevant equations in (11) and (14), I respectively use the two mass parameters $M_{Z'_{\mu}}$ and $M_{\nu_{4}}$ to take the place of them. The $M_{Z'_{\mu}}$ value is bounded by both the relation $M_{Z'_{\mu}}\geqslant2g_{1}v_{\phi}$ and the experimental limits, while the value of $M_{\nu_{4}}$ is determined by fitting the current observations of the relic abundance of the cold dark matter, namely $\Omega h^{2}=0.112$ \cite{34}. A set of reasonable values of the two mass parameters are chosen as
\ba
 M_{Z'_{\mu}}=2\:\mathrm{TeV},\hspace{0.5cm} M_{\nu_{4}}=788\:\mathrm{GeV}.
\ea
 According to (11), (18) and (19), the gauge mixing angle $sin\theta_{g}$ and the $\nu_{4L}$ relic abundance are calculated as follows
\ba
 sin\theta_{g}=0.522\,,\hspace{0.5cm} \Omega h^{2}=0.112\,.
\ea
 The above results are very desired.
\begin{figure}
 \centering
 \includegraphics[totalheight=9cm]{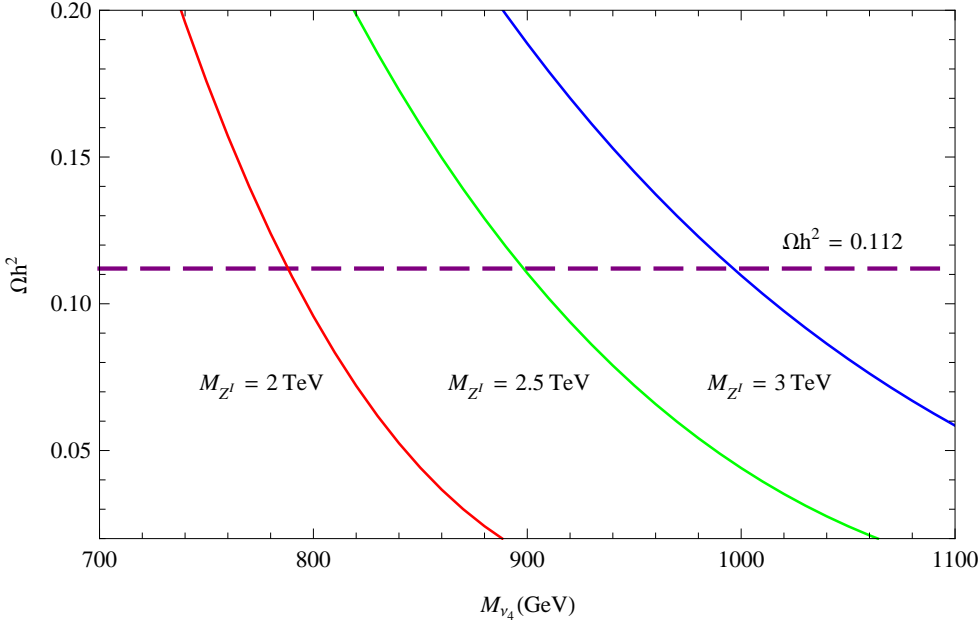}
 \caption{The graphs of $\Omega h^{2}$ versus $M_{\nu_{4}}$ for $M_{Z'_{\mu}}=[2,\,2.5,\,3]$ TeV, the red curve corresponds to the case in (29)\,.}
\end{figure}
 In the figure (4), I draw the graphs of $\Omega h^{2}$ versus $M_{\nu_{4}}$ for the three values of $M_{Z'_{\mu}}=[2,\,2.5,\,3]$ TeV, the red curve exactly corresponds to the case in (29). A smaller value of $M_{Z'_{\mu}}$ can not satisfy the inequality condition and the experimental limits, whereas a larger $M_{Z'_{\mu}}$ can lead to a very small $sin\theta_{g}$, this is also unacceptable. Based on a overall consideration, a reasonable region of $M_{Z'_{\mu}}$ should be about $1.5-3.5$ TeV, therefore, the corresponding values of $\nu_{4L}$ should lie in $700-1100$ GeV. Because $\nu_{4L}$ has no any direct interactions with the SM particles, however, it will be difficult for searching it in future experiments.

 Finally, I analyze the leptogenesis. The four parameters, $\lambda_{4},\,c_{1e},\,c_{2e},\,M_{S}$, have to be added to fulfil the calculations. The reasonable values of $\lambda_{4},\,c_{1e},\,c_{2e}$ are all around $0.1$. Since the role of $c_{2e}$ is relatively insignificant, I directly take $c_{2e}=c_{1e}$ for simplicity. The value of $M_{S}$ is determined by fitting the baryon asymmetry. In addition, the order of magnitude of the parameter $y^{e}_{4}$ is $\leqslant 10^{-9}$. In short, these parameters are chosen as the following values,
\ba
  y^{e}_{4}\leqslant 10^{-9},\hspace{0.5cm} c_{1e}=c_{2e}=0.1\,,\hspace{0.5cm} \lambda_{4}=0.1\,,\hspace{0.5cm} M_{S}=907.3\:\mathrm{GeV}.
\ea
 By use of (20)--(22), the ratio of the $S^{\mp}$ decay width to the Hubble expansion rate as well as the baryon asymmetry are calculated as follows
\ba
 \frac{\Gamma(S^{-}\rightarrow e^{-}_{i}+\overline{\nu_{4}})}{H(T=M_{S})}\leqslant 1.9\times10^{-8}\,,\hspace{0.5cm}
 \eta_{B}=6.15\times10^{-10}.
\ea
 It can be seen from the above results that the $S^{\mp}$ decay is indeed out-of-equilibrium, and also $\eta_{B}$ is precisely in agreement with the current data of the baryon asymmetry \cite{35}. Of course, $M_{S}\sim1$ TeV is also in accordance with the previous expectation. It should especially be emphasized that the baryon asymmetry can still be generated though the lepton $CP$ violation vanishing. The reason for this is that the fourth generation charged lepton and neutrino are bound to inhabit in the universe exactly as the model descriptions.
\begin{figure}
 \centering
 \includegraphics[totalheight=9cm]{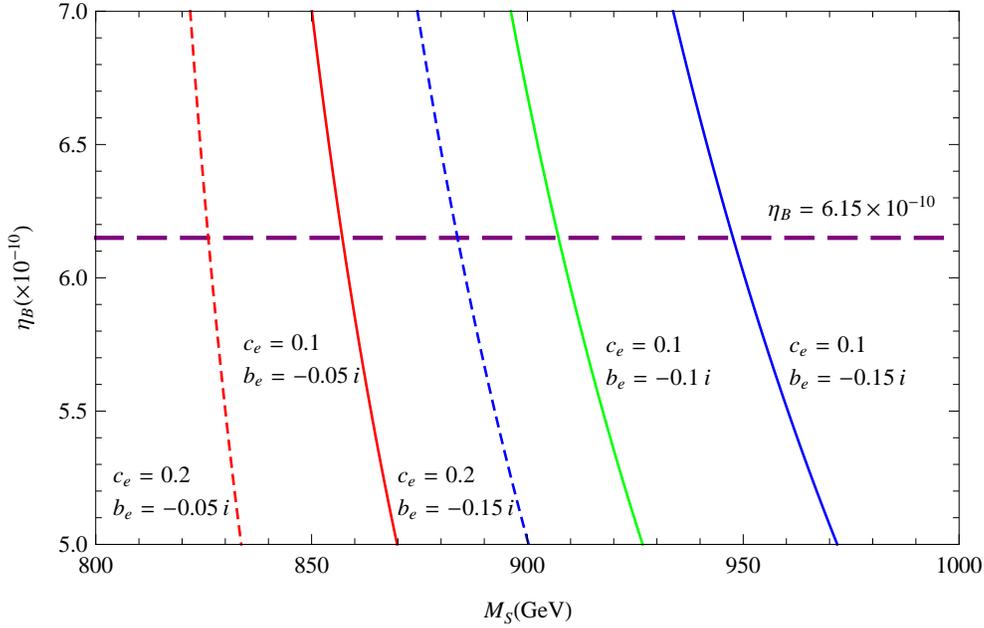}
 \caption{The graphs of the baryon asymmetry subjecting to $M_{S}$ for the five set of values of the parameters $[c_{1e}=c_{2e}=c_{e},b_{e}]$ when the other parameters are fixed in the context, the green curve corresponds to the case in (26) and (31)\,.}
\end{figure}
 The baryon asymmetry subjecting to $M_{S}$ is drawn in the figure (5), in which the parameters $[c_{1e}=c_{2e}=c_{e},b_{e}]$ are chosen as the five set of values, while the other parameters are fixed in the context, the green curve exactly corresponds to the case in (26) and (31). It can be inferred from the figure that the reasonable area of $M_{S}$ should be about $800-1000$ GeV. Therefore, there is a great chance to find the $S^{\mp}$ particle in the future.

 To summarize all kinds of the above numerical results, the model excellently and accurately fits all the current experimental data of the fermion masses and flavor mixings, and the cold dark matter and baryon asymmetry. All of the current measured values are correctly reproduced, while all of the non-detected values are finely predicted in the experimental limits. All the results are naturally produced without any fine tuning. In particular, the model gives a number of interesting predictions.
 A mass spectrum of all kind of the model particles is summarized as follows
\begin{alignat}{1}
 & M_{\nu}\sim 0.01\,\mathrm{eV} \ll M_{SM particles}\sim (0.001-100)\,\mathrm{GeV} \nonumber\\
 & < M_{\nu_{4}}\sim 800\,\mathrm{GeV} < (M_{\phi^{0}},M_{S})\sim 1\,\mathrm{TeV} \nonumber\\
 & < (M_{Z'_{\mu}},M_{e_{4}},M_{u_{4}},M_{d_{4}})\sim (2-3)\,\mathrm{TeV} \ll M_{\Delta}\sim 10^{5}\,\mathrm{TeV}.
\end{alignat}

 In the end, I give some methods how to test the model at the LHC. On the basis of the model interactions, the figure (6) draws some optimum mechanisms of producing the model non-SM particles by the proton-proton collisions.
\begin{figure}
 \centering
 \includegraphics[totalheight=4.5cm]{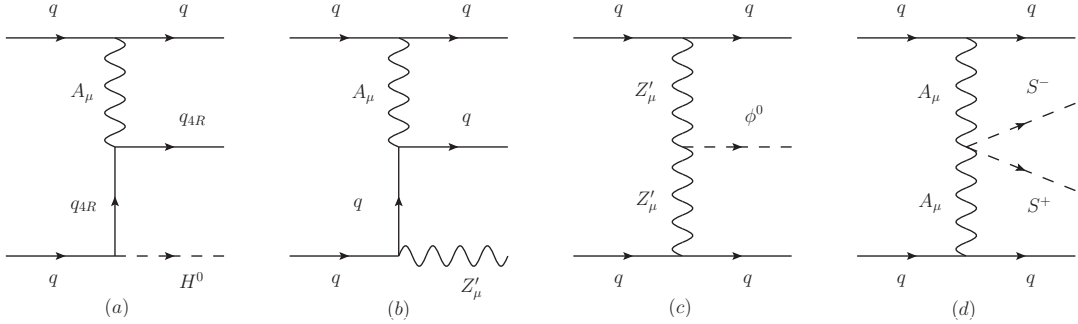}
 \caption{The diagrams of producing some non-SM particles of the model at the LHC, in which $A_{\mu}$ denotes photon.}
\end{figure}
 (a) diagram can produce the fourth generation quarks if the colliding energy is enough effective, moreover, it is also a new channel of producing the SM Higgs boson $H^{0}$. (b) and (c) respectively illustrate the productions of the heavy gauge boson $Z'_{\mu}$ and scalar boson $\phi^{0}$. The charged scalar boson pair $S^{\mp}$ can be produced by the (d) process. The fourth generation charged lepton and the cold dark matter neutrino are both not able to be produced in similar ways, but they can be found in the decay products of $Z'_{\mu},\,\phi^{0},\,S^{\mp}$, namely $Z'_{\mu}\rightarrow e_{4}\overline{e_{4}}$, $Z'_{\mu}\rightarrow \nu_{4}\overline{\nu_{4}}$, $\phi^{0}\rightarrow \nu_{4}\nu_{4}$, $S^{\mp}\rightarrow e^{\mp}_{4}\nu_{4}$. The loss of energy in the decay processes should be regarded as a definitive signal of the cold dark matter neutrino $\nu_{4}$. In particular, the (d) process produces a pair of $S^{\mp}$, in succession, $S^{\mp}$ decay into $e^{\mp}$ and $\nu_{4}$, the final sate leptons will eventually generate a matter-antimatter asymmetry. In other word, the TeV scale collider can experimentally produce the matter-antimatter asymmetry. In short, the model is quite feasible to be tested at the LHC as long as the luminance and running time are enough large. Undoubtedly, the best efficient methods to test the model are of course by the lepton-antilepton collisions at the ILC. To save space, here I do not go into details. A full discussion of the test of the model is planed in another paper.

\vspace{1cm}
 \noindent\textbf{VI. Conclusions}

\vspace{0.3cm}
 In the paper, I have suggested a practical and feasible particle model to account for the fermion flavor puzzle, the cold dark matter, and the matter-antimatter asymmetry. The model has the TeV scale $U(1)_{B-L}$ symmetry. It contains several new particles with the TeV scale masses, for example, the gauge boson $Z'_{\mu}$, the scalar bosons $\phi^{0}$ and $S^{\mp}$, and the fourth generation quarks and leptons, in which the fourth generation neutrino is exactly the cold dark matter. In addition, the model has also the very heavy scalar boson $\Delta$ whose mass is $\sim10^{5}$ TeV, by the new form seesaw $\Delta$ develops only the tiny VEV. This is the essential source of the tiny masses of the light neutrinos. These non-SM particles play key roles in the model. The model can not only excellently explain the fermion masses and flavor mixings, but also elegantly accommodate the cold dark matter and leptogenesis at TeV scale, moreover, it gives some interesting predictions. The theory can perfectly integrate three party of the flavor physics, the cold dark matter and the matter-antimatter asymmetry, therefore, it is quite deserved to be tested in future experiments on the ground and in the sky. Finally, I believe that the new non-SM particles, including the cold dark matter neutrino $\nu_{4}$, are possible to be discovered at the LHC and ILC in the future. In a word, all kinds of the experiments toward these goals will not only provide us more information about particle physics, but also help us understand the mysteries of the universe.

\vspace{1cm}
 \noindent\textbf{Acknowledgments}

\vspace{0.3cm}
 I would like to thank my wife for large helps. I also thank Ayres Freitas for his helpful comments. This research is supported by chinese universities scientific fund.

\end{document}